\documentclass[aps,prb,twocolumn,superscriptaddress,showpacs]{revtex4}

\usepackage{graphicx}
\usepackage{dcolumn}
\usepackage{bm}

\begin{document}

\title{Large magnetothermal conductivity in
GdBaCo$_2$O$_{5+x}$ single crystals}

\author{X. F. Sun}
\email[]{xfsun@ustc.edu.cn}

\affiliation{Hefei National Laboratory for Physical Sciences at
the Microscale, University of Science and Technology of China,
Hefei, Anhui 230026, P. R. China}

\affiliation{Central Research Institute of Electric Power
Industry, Komae, Tokyo 201-8511, Japan}

\author{A. A. Taskin}
\affiliation{Central Research Institute of Electric Power
Industry, Komae, Tokyo 201-8511, Japan}

\author{X. Zhao}
\affiliation{Department of Astronomy and Applied Physics,
University of Science and Technology of China, Hefei, Anhui
230026, P. R. China}

\author{A. N. Lavrov}
\affiliation{Institute of Inorganic Chemistry, Lavrentyeva-3,
Novosibirsk 630090, Russia}

\author{Yoichi Ando}
\affiliation{Institute of Scientific and Industrial Research,
Osaka University, Ibaraki, Osaka 567-0047, Japan}

\date{\today}

\begin{abstract}

To study the effects of paramagnetic spins on phonons, both the
in-plane and the $c$-axis heat transport of GdBaCo$_2$O$_{5+x}$
(GBCO) single crystals are measured at low temperature down to
0.36 K and in magnetic field up to 16 T. It is found that the
phonon heat transport is very strongly affected by the magnetic
field and nearly 5 times increase of the thermal conductivity in
several Tesla field is observed at 0.36 K. It appears that phonons
are resonantly scattered by paramagnetic spins in zero field and
the application of magnetic field removes such strong scattering,
but the detailed mechanism is to be elucidated.

\end{abstract}

\pacs{66.70.+f, 72.20.-i}

\maketitle

\section{Introduction}

In magnetic materials, coupling of spins with other elementary
excitations is a matter of fundamental importance. In particular,
the spin-phonon interaction causing unusual physical properties in
correlated oxides has been an issue of significant interest both
by its own right and because it provides a useful means to gain
insight into frustrated magnetism as well as multiferroics.
\cite{Kimura, Sharma, Sushkov, Hemberger, Gupta} The
low-temperature thermal conductivity is one of the most effective
probes for studying the spin-phonon coupling,\cite{Berman}
especially in insulators where magnetic excitations along with
phonons are the main heat carriers. Even if the long-range
magnetic order is not established in a crystal or the 
contribution of magnons to the thermal conductivity is limited at 
low temperatures, the presence of localized spins still can play 
an important role in the heat transport as was observed long time 
ago in some ionic compounds doped with magnetic 
impurities.\cite{Berman, Walton, Fox} However, in strongly 
correlated materials, the low-$T$ behavior of thermal 
conductivity can be so complicated that it is difficult to 
selectively study the effect of spin-phonon interaction, 
especially when the electron transport, which can also depend on 
magnetic field, is contributing to the thermal conductivity. In 
order to gain insight into the impact of spin-phonon coupling and 
its underlying mechanism, one needs a material in which the 
spin-phonon coupling can be treated independently.

For this purpose, recently synthesized layered cobaltite
GdBaCo$_2$O$_{5+x}$ (GBCO), which has attracted a great deal of
attention due to its rich phase diagram and fascinating physical
properties (See Ref. \onlinecite{Taskin1} and references therein),
would be an ideal system for studying the spin-phonon coupling.
This compound is a good electrical insulator at low temperatures
for a wide range of oxygen content, and hence there is no need to
consider the electronic heat conduction. Although GBCO shows a
long-range magnetic ordering of Co spins, their Ising-like spin
anisotropy prevents the low-energy magnon excitations, while the
Gd ions remain paramagnetic and behave as almost free spins down
to very low temperatures. Furthermore, a lattice disorder in GBCO
can be easily introduced by merely changing its oxygen content.

In this work, we perform detailed studies of the low-$T$ heat
transport of GBCO single crystals with various oxygen contents. It
is found that both the in-plane and the $c$-axis thermal
conductivities ($\kappa_{ab}$ and $\kappa_c$) show extremely
strong magnetic-field dependence at low temperatures. The
$\kappa(H)$ isotherms at 1--2 K show a dip-like feature at low
fields where the heat conduction is reduced by as much as 50\%,
while at subkelvin temperatures the magnetic field induces a
step-like enhancement of $\kappa$ by nearly 5 times. The
magnetic-field dependences of the thermal conductivity indicate
that phonons are strongly scattered by the Gd spins that are
susceptible to the Zeeman effect in magnetic fields. The large
magnetothermal effect illuminates the important role of free 
spins in the heat transport of novel oxide materials.

\section{Experiments}

High-quality GdBaCo$_2$O$_{5+x}$ single crystals are grown using 
a floating-zone technique\cite{Taskin1, Taskin2, Taskin3} and 
carefully annealed to tune their oxygen contents to $x=0.50$, 
0.30 and 0.00. The thermal conductivity along both the $ab$ plane 
and the $c$ axis is measured from 0.36 to 5 K in a $^3$He 
refrigerator by using a ``one heater, two thermometers" technique 
and from 5 to 300 K in a $^4$He cryostat by using the 
Chromel/Constantan thermocouple. Details of the measurements of 
the temperature and magnetic-field dependence of $\kappa$ were 
described in previous publications.\cite{Sun_PLCO, Sun_LSCO, 
Sun_PLCCO} All the parallelepiped crystals are precisely cut 
along three crystallographic axes with an error of less than 
1$^\circ$. The heat capacity is measured by the relaxation method 
in the temperature range from 2 to 300 K using a commercial 
physical properties measurement system (PPMS, Quantum Design).

\section{Results and discussions}

\begin{figure}
\includegraphics[clip,width=8.5cm]{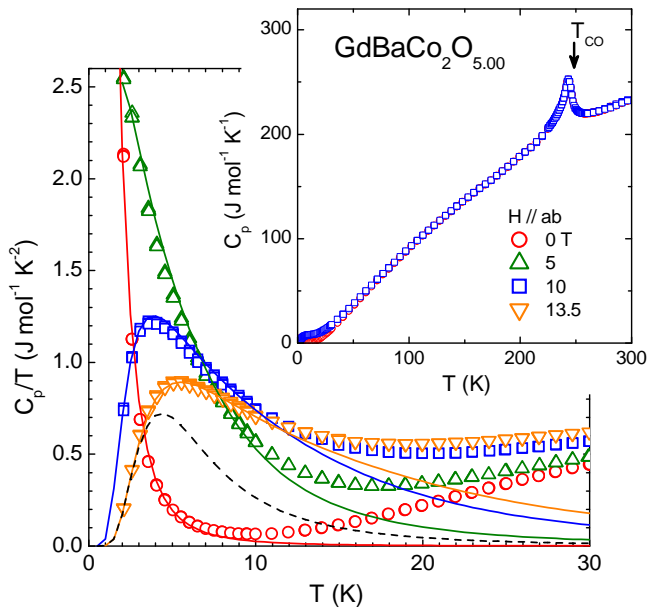}
\caption{(color online) The low-temperature specific heat of 
GdBaCo$_2$O$_{5.00}$ in magnetic fields of 0, 5, 10, and 13.5 T 
applied along the $ab$ plane. Solid lines show calculated 
Schottky contributions of paramagnetic Gd$^{3+}$ ions assuming 
$S$ = 7/2 (see text). The dashed line indicates the Schottky 
contribution of paramagnetic ions with $S$ = 1/2 at $H$ = 13.5 T. 
The inset shows the heat capacity in the temperature range of 
2--300 K measured at $H$ = 0 and 10 T. The arrow indicates the 
charge ordering transition.}
\end{figure}

Before presenting the heat transport results of GBCO crystals, we 
show in Fig. 1 the representative heat capacity data, which prove 
to be very informative regarding the magnetic excitations. In 
GBCO, regardless of the oxygen content, the spins of cobalt ions 
undergo magnetic ordering at rather high 
temperatures,\cite{Taskin1, Taskin2} but Gd ions, in contrast, 
remain paramagnetic down to very low temperatures as suggested by 
magnetization measurements.\cite{Taskin1} In order to confirm 
this, we have measured the heat capacity of an $x$ = 0.00 single 
crystal in different magnetic fields. As can be seen in the inset 
of Fig. 1, the $C_p(T)$ curve is insensitive to the magnetic 
field, except for the low-$T$ part: the peak around 248 K, which 
corresponds to the charge ordering transition at this 
temperature,\cite{Suard} is not affected by applying a high 
magnetic field either. At low temperatures, on the other hand, 
the specific heat $C_p/T$ shows a pronounced Schottky anomaly 
with a maximum that shifts to a higher temperature with 
increasing magnetic field. All the data measured in $H$ = 0, 5, 
10, and 13.5 T are well fitted by a simple Schottky contribution 
of paramagnetic Gd$^{3+}$ ions with a spin value of 7/2, which is 
determined by the area under the peak (the Schottky contribution 
of paramagnetic ions with $S$ = 1/2 at $H$ = 13.5 T is shown for 
comparison), and the $g$ factor of 1.64 $\pm$0.04, which is 
determined by the peak position. We should note also that there 
is a finite splitting of the Gd spin levels even in the absence 
of an external magnetic field that can be a result of a weak 
magnetic interaction of Gd ions with each other or with the 
cobalt sublattices.  

\begin{figure}
\includegraphics[clip,width=8.5cm]{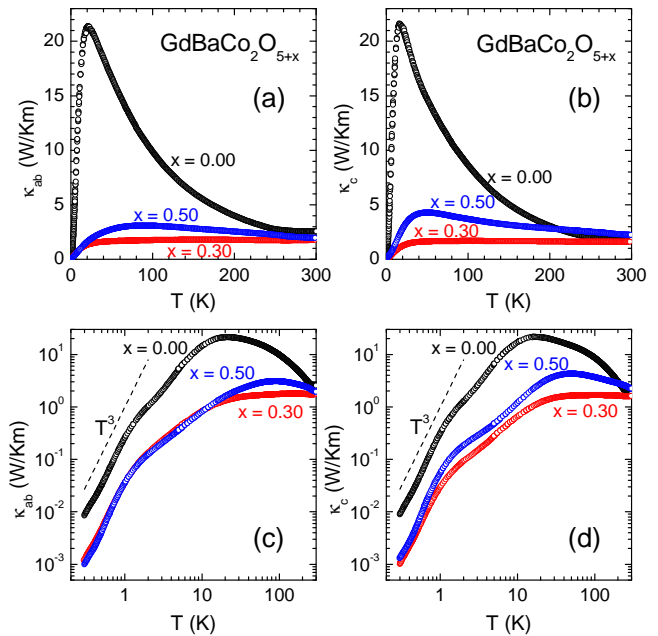}
\caption{(color online) Temperature dependences of the thermal
conductivity along the $ab$ plane and the $c$ axis in zero
magnetic field for GdBaCo$_2$O$_{5+x}$ single crystals with
various oxygen content. The data are displayed in both linear and
logarithmic plots. The dashed lines indicate the $T^3$
dependence.}
\end{figure}

Figure 2 shows the temperature dependences of $\kappa_{ab}$ and
$\kappa_c$ of GdBaCo$_2$O$_{5+x}$ single crystals in zero field.
First of all, as we already pointed out, the insulating ground
state and the Ising-like spin anisotropy of GBCO leave little room
for carriers other than phonons to contribute to the heat
transport, at least at low temperatures. As can be seen in Fig. 2,
the $x = 0.00$ crystals show a large phonon peak at $\sim$ 20 K,
indicating the negligible imperfection of the crystal lattice. In
contrast, for $x = 0.30$ and 0.50, the phonon conductivity is
dramatically weakened at low temperatures and the phonon peak is
wiped out almost completely, which is not surprising since the
oxygen non-stoichiometry brings strong lattice disorder. Between
the latter two compositions, the $x = 0.50$ crystals have better
heat conductivity, presumably due to the oxygen ions being ordered
into chains at this oxygen content.\cite{Taskin1, Taskin2} The
worst heat conduction is observed in $x = 0.30$ crystals, which
have the largest degree of oxygen-induced lattice 
disorder.\cite{Taskin1}

The temperature dependences of $\kappa$ in Fig. 2 are actually
more complicated than those in conventional insulators. At high
temperatures, for example, the $\kappa(T)$ data for the good
phonon conductor $x = 0.00$ cannot be described by the usual
phonon Umklapp scattering,\cite{Berman, note} which suggests the
existence of a more complicated scattering mechanism or some other
kind of heat carriers. In fact, the specific heat data in Fig. 1 
indicate that above $\sim$100 K, magnon excitations of the Co 
spin lattice become important and may affect the high-$T$ heat 
transport properties. On the opposite side, the lowest-$T$ 
behavior of the heat conduction also seems to differ from the 
boundary scattering limit of phonons expected for conventional 
insulators,\cite{Berman} in which case the magnitude of the 
phonon thermal conductivity should depend only on the sample 
size, and not on impurities, defects, or oxygen concentration 
(the dispersion of acoustic phonons and the sound velocity do not 
change much upon changing the oxygen content). Although both 
$\kappa_{ab}$ and $\kappa_c$ below 1 K are rather close to $\sim 
T^3$, the heat conductivity of the $x$ = 0.00 sample is nearly 
one order of magnitude larger than that of $x = 0.30$ and 0.50 
samples. Note that the $\kappa_c$ data for different oxygen 
contents were collected on one and the same sample that was 
annealed each time to obtain the required $x$; the samples for 
$\kappa_{ab}$ differ in size by less than 20\%. In addition, the 
low-$T$ parts of the $\kappa(T)$ data are not smooth: regardless 
of the oxygen content, all the crystals exhibit clear wiggles in 
their $\kappa(T)$ curves below 2 K, which suggests the existence 
of some kind of resonant phonon scattering.\cite{Berman}

\begin{figure}
\includegraphics[clip,width=8.5cm]{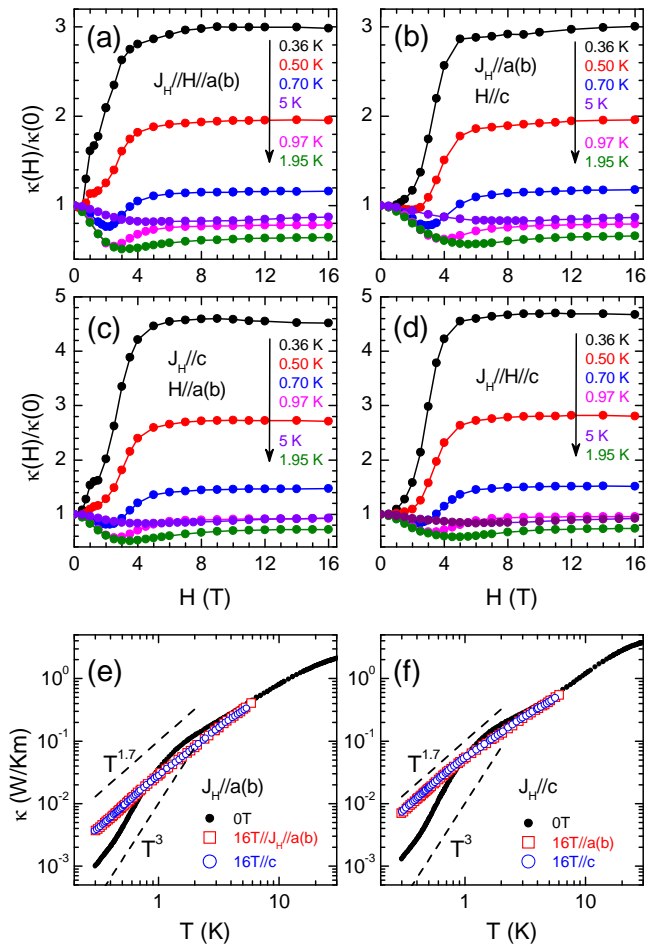}
\caption{(color online) (a--d) Magnetic-field dependences of the
low-$T$ thermal conductivity of the GdBaCo$_2$O$_{5+x}$ single
crystal with $x=0.50$. Directions of the heat current ($J_H$) and
the magnetic field ($H$) are shown in each panel. (e,f)
Temperature dependences of $\kappa_{ab}$ and $\kappa_c$ in zero
and 16 T magnetic fields. The dashed lines indicate the $T^3$ and
$T^{1.7}$ dependences.}
\end{figure}

Upon applying the magnetic field at low temperatures, we have
found remarkably strong changes in the heat conductivity of all
crystals. For instance, at the lowest temperature 0.36 K, both
$\kappa_{ab}$ and $\kappa_c$ of the parent compound $x=0.50$ show
a step-like enhancement by several times with increasing field and
saturates above $\sim$5 T [Figs. 3(a-d)]. At higher temperatures,
a dip-like feature appears at low field, while the subsequent
enhancement at higher field gradually weakens with increasing
temperature. At the ``high'' temperature of 5 K, the
magnetic-field dependence is already rather weak and only a broad
and shallow dip remains. Here, the most impressive result is the
magnitude of the field-induced changes in $\kappa$: {\it the
low-field suppression can be as large as a factor of 2, and the
high-field enhancement can be nearly 400\%.} While the magnitude
of the high-field enhancement in $\kappa$ depends on the direction
of the heat current [Figs. 3(a-f)], it is surprisingly insensitive
to the field direction, which confirms that magnon excitations in
the long-range ordered spins cannot be relevant to the low-$T$
heat transport.

A clue given by the dip-like feature in $\kappa(H)$ is that the 
low-$T$ heat transport may be governed by the phonon scattering 
by free spins,\cite{Berman, Walton, Fox} as we previously 
observed in Pr$_{1.3}$La$_{0.7}$CuO$_4$ (PLCO).\cite{Sun_PLCO} 
Qualitatively, the $\kappa(H)$ behavior shown in Fig. 3 can be 
well understood in terms of magnetic scattering of phonons: (i) 
In zero field, the phonons are presumably scattered by the spins 
of Gd ions that behave like $S=7/2$ free spins judging from the 
low-$T$ magnetization and specific heat data.\cite{Taskin1, 
Taskin2} To be capable of scattering phonons, the Gd spins should 
not be completely free, but the spin states should be slightly 
splitted, which has been confirmed by the heat capacity data. 
(ii) In magnetic fields, the energy splitting of the spin states 
is increased by the Zeeman effect. The phonon scattering off 
these spins is most effective in suppressing the heat transport 
when the energy splitting is equal to $\sim 3.8k_BT$, where the 
phonon conductivity spectrum (defined below) peaks; therefore, 
the spin-phonon scattering generates a dip-like feature in 
$\kappa(H)$ at this energy and the dip position shifts to higher 
fields with increasing temperature.\cite{Berman} (iii) In the 
high-field limit, the spin energy splitting becomes too large to 
exchange energy with phonons and the spin-phonon scattering is 
completely quenched, enhancing $\kappa$ above its zero-field 
value.

\begin{figure}
\includegraphics[clip,width=8.5cm]{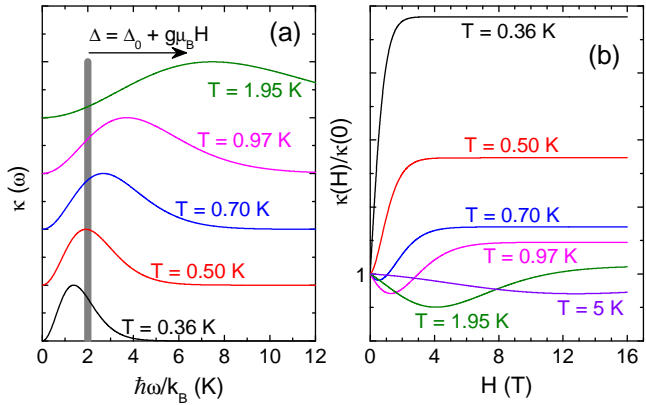}
\caption{(color online) (a) Schematic picture of the phonon
conductivity spectrum $\kappa (\omega)$ at several temperatures
(the curves are normalized by their peak values and shifted
vertically for clarity). The phonons can be strongly scattered by
magnetic ions if their energy is close to the energy splitting of
the spin states, which gives rise to a ``resonant scattering band"
centered at $\Delta = \Delta_0 + g\mu_BH$, where $\Delta_0$ is the
zero-field splitting and $g\mu_BH$ is the Zeeman splitting. (b)
Simulated $\kappa(H)$ by assuming that the phonons within the
resonant scattering band do not carry heat.}
\end{figure}

To illustrate the above picture, in Fig. 4 we show schematically
the relation between the energy splitting of spin states and the
peak in the phonon conductivity spectrum from the Debye model. The
``phonon conductivity spectrum" $\kappa (\omega)$ represents the
contribution to thermal conductivity from the phonons with an
energy $\hbar \omega$ and is proportional to $\omega^4 \exp(\hbar
\omega/k_B T)/T^2 [\exp(\hbar \omega/k_B T)-1]^2$ which peaks at
energy $\hbar \omega \sim 3.8 k_BT$.\cite{Berman} Although Gd 
ions have the spin state $S=7/2$ that splits into eight energy 
states in magnetic fields, allowing multiple transitions, for 
simplicity we take just a single band, corresponding to the spin 
transition with the lowest energy, $\Delta m=1$. The energy of 
this transition changes with the magnetic field as $\Delta = 
\Delta_0 + g\mu_BH$, and the phonons in a certain energy range 
around this $\Delta$ are considered to be resonantly scattered 
upon the spin transition. Figure 3(b) shows simulated $\kappa(H)$ 
behaviors, where one can see the variations of $\kappa$ as the 
position of the resonant scattering band is changed by the 
magnetic field. Upon calculating these $\kappa(H)$, we simply 
assume that those phonons whose energy lies in a resonant 
scattering band centered around $\Delta$ cannot contribute to the 
heat transport, and the rest of the phonons have the same 
relaxation time (which is the case at very low 
temperature).\cite{Hetzler} It is reassuring that such a crude 
picture with only two adjustable parameters, $\Delta_0$ 
($\approx$2 K) and the width of the resonant band, can capture 
most of the qualitative experimental features of $\kappa(H)$. 
Indeed, both the step-like enhancement and the dip-like feature 
are essentially reproduced.

Nonetheless, there is an obvious difficulty in the above simple
picture; namely, the magnitude of the observed changes in
$\kappa(H)$ requires the resonance scattering band to be extremely
wide; indeed, the observed five-fold increase in $\kappa$ implies
that at least 80\% of phonon conductivity spectrum are affected by
the spin scattering. Apparently, including other resonance bands
related to higher energy spin transitions of Gd ions with $\Delta
m>1$ does not help, since the Zeeman splitting $\Delta m g\mu_BH$
should quench these transitions at several times lower fields than
we observe in the experiment. A possible explanation for the
extremely wide scattering band is the local fluctuation of the
zero-field splitting $\Delta_0$, which would be quite natural if
$\Delta_0$ is determined by the dipole-dipole interactions within
the disordered Gd subsystem. Depending on the local environment,
Gd ions feel different local fields and thus should exhibit a
continuous spectrum of spin-transition energies, rather than a
single resonance line.\cite{Glazkov}

\begin{figure}
\includegraphics[clip,width=8.5cm]{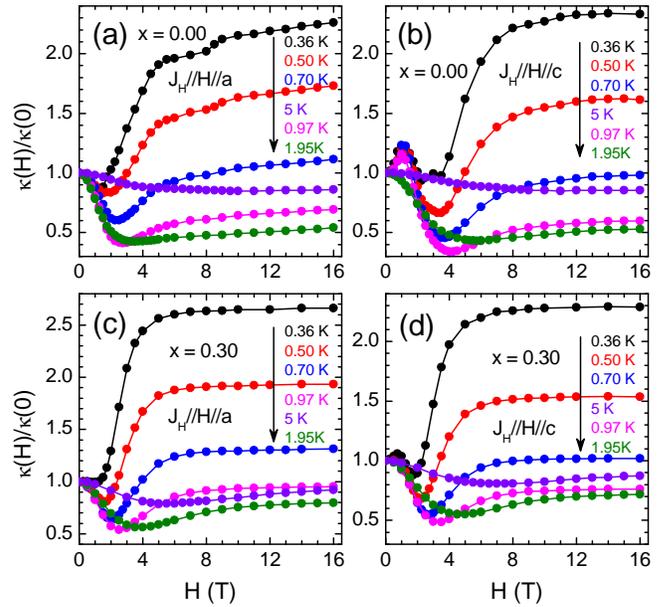}
\caption{(color online) Representative data for the magnetic-field
dependences of the low-$T$ thermal conductivity of
GdBaCo$_2$O$_{5+x}$ single crystals with $x$ = 0.00 and 0.30.}
\end{figure}

We have done the same experiments on $x$ = 0.30 and 0.00 crystals 
and show the representative $\kappa(H)$ isotherms in Fig. 5. 
Obviously, the main features, including both the high-field 
step-like enhancement and low-field dip-like feature, are the 
same as those for $x$ = 0.50 crystals. The essential similarity 
of the $\kappa(H)$ behaviors for different oxygen contents seems 
reasonable if the phonon scattering by Gd moments is playing the 
key role, since the oxygen content apparently has no major impact 
on the magnetic state of the Gd ions. This result clearly 
indicates the separable effects of free spins and 
impurities/defects on the low-$T$ heat transport of GBCO. It is 
worth noting that the quantitative $x$ dependence of 
$\kappa(H)/\kappa(0)$ is rather complicated, namely, the 
magnitude of $\kappa(H)/\kappa(0)$ is slightly enhanced upon 
increasing $x$ from 0.00 to 0.30, while it becomes much more 
pronounced upon further increasing $x$ to 0.50. Obviously, for 
quantitative description of the all experimental observations, 
not only the spin-phonon scattering but the oxygen-induced 
lattice disorder must be taken into consideration. 

Although the detailed mechanism of the large magnetothermal 
effect in GBCO calls for further investigation, one can already 
capture some important information from the above experimental 
results. First, the phonon heat transport can be strongly 
dependent on the magnetic field even at very low temperature for 
the materials containing paramagnetic moments, like high-$T_c$ 
cuprates\cite{Sun_PLCO} and other strongly correlated compounds. 
Second, if the paramagnetic moments have a small enough splitting 
between the spin states, the spin-phonon scattering can survive 
down to very low temperature and prevent phonons from entering 
the boundary scattering limit. This makes the data analysis of 
the low-$T$ thermal conductivity very difficult, especially when 
the boundary scattering limit is required for separating the 
electron and phonon terms.\cite{Taillefer, Sutherland_PRB, 
Sun_logT, Sun_nonuniversal}

\section{Summary}

In summary, a surprisingly strong magnetic field dependence of 
thermal conductivity is observed in GBCO single crystals down to 
very low temperatures. The main features of the magnetothermal 
conductivity, {\it i.e.}, a high-field enhancement and a 
low-field dip, can be well understood in terms of the phonon 
scattering by the nearly free Gd spins. The present finding 
demonstrates the potentially significant role of the spin-phonon 
coupling in correlated oxides.

\begin{acknowledgments}

We thank J. Takeya for helpful discussions. This work was
supported by the National Natural Science Foundation of China
(50421201 and 10774137), the National Basic Research Program of
China (2006CB922005), and KAKENHI 16340112 and 19674002.
\end{acknowledgments}


\begin{thebibliography}{}
\bibitem{note}
The high-$T$ $\kappa(T)$ data cannot be fitted well to the Debye
formula for the phononic thermal conductivity.\cite{Berman}
\end{thebibliography}
\end{document}